\documentclass[conference]{IEEEtran}
\IEEEoverridecommandlockouts
\usepackage[top=0.75in, bottom=1.1in, left=0.625in, right=0.625in]{geometry}

\usepackage{cite}
\usepackage{amsmath,amssymb,amsfonts}
\usepackage{algorithmic}
\usepackage{graphicx}
\usepackage{subcaption}  
\usepackage{textcomp}
\usepackage{xcolor}
\usepackage{flushend}
\def\BibTeX{{\rm B\kern-.05em{\sc i\kern-.025em b}\kern-.08em
    T\kern-.1667em\lower.7ex\hbox{E}\kern-.125emX}}
\begin{document}

\setlength{\columnsep}{0.24in}

\title{Secure Time-Modulated Intelligent Reflecting Surface via Generative Flow Networks
\thanks{This work was supported by ARO grant W911NF2320103 and NSF grant ECCS-2320568.}
}

\author{\IEEEauthorblockN{Zhihao Tao}
\IEEEauthorblockA{\textit{Electrical and Computer Engineering} \\
\textit{Rutgers, the State University of New Jersey}\\
New Brunswick, USA \\
zt118@scarletmail.rutgers.edu}
\and
\IEEEauthorblockN{Athina P. Petropulu}
\IEEEauthorblockA{\textit{Electrical and Computer Engineering} \\
\textit{Rutgers, the State University of New Jersey}\\
New Brunswick, USA \\
athinap@soe.rutgers.edu}
}

\maketitle

\begin{abstract}
We propose a novel directional modulation (DM) design for OFDM transmitters aided by a time-modulated intelligent reflecting surface (TM-IRS). The TM-IRS is configured to preserve the integrity of transmitted signals toward multiple legitimate users while scrambling the signal in all other directions.
Existing TM-IRS design methods typically target a single user direction and follow predefined rule-based procedures, making them unsuitable for multi-user scenarios. 
Here, we propose a generative AI-based approach
to design good sets of  TM-IRS parameters out of a set of all possible quantized ranges of parameters.
 The design objective is to maximize the sum rate across the authorized directions.
We model the TM-IRS parameter selection as a deterministic Markov decision process (MDP), where each terminal state corresponds to a specific configuration of TM-IRS parameters. GFlowNets are employed to learn a stochastic policy that samples TM-IRS parameter sets with probability proportional to their associated sum rate reward. 
Experimental results demonstrate that the proposed method 
effectively enhances the security of the TM-IRS-aided OFDM systems with multi-users. Also, despite the vast size of the TM-IRS configuration space, the GFlowNet is able to converge after training on fewer than 0.000001\% of all possible configurations, demonstrating remarkable efficiency compared to exhaustive combinatorial search.
Implementation code is available at 
\text{https://github.com/ZhihaoTao/GFN4TM-RIS}
to facilitate reproducibility.
\end{abstract}

\begin{IEEEkeywords}
Intelligent reflecting surface, time modulation, physical layer security, OFDM, GFlowNets.
\end{IEEEkeywords}

\section{Introduction}
The broadcast nature of wireless propagation exposes confidential data to interception unless special precautions are taken. Physical layer security (PLS) exploits physical characteristics of the wireless medium, such as channel fading, noise, interference, and spatial diversity, to complement—or, in some circumstances, replace—higher-layer cryptographic techniques~\cite{Shannon1949Comm,Wyner1975Wire,poor2017wireless}. Among the many PLS mechanisms proposed, directional modulation (DM) has attracted particular interest because it embeds information in the spatial signature of the transmitted waveform: a receiver aligned with the intended steering direction observes an undistorted constellation, whereas other directions see a scrambled one~\cite{daly2009dire,qiu2023decomposed,tao2024tma}. Compared with other PLS approaches such as secrecy rate maximization \cite{lv2015secrecy, gong2016millimeter} or artificial-noise injection \cite{zhang2019AN,wang2017AN}, DM can offer comparable secrecy in a more cost-effective and energy-efficient manner \cite{su2022secure}.

DM implementations have been proposed for fully digital or hybrid beamforming architectures with multiple radio-frequency (RF) chains and fine-grained phase control at each antenna element or each transmitted symbol \cite{li2019performance, Ottersten2016, Alodeh2016DM, su2022secure}. 
A cost-effective single-RF-chain alternative is to apply time modulation to a phased array driven by orthogonal frequency-division multiplexing (OFDM) signals~\cite{tvt2019time, tao2025twc}. By using single-pole-single-throw (SPST) switches to periodically connect   and disconnect antennas to the RF chain, a time-modulated array (TMA) generates controllable harmonics whose periods are aligned with the OFDM symbol duration. As a result, each subcarrier of the transmitted OFDM signal  carries a weighted mixture of all original symbols, where the mixing coefficients depend on the TMA parameters, i.e., connection times, or on states, and on state durations. This mixing represents scrambling of the transmitted symbols in all directions.
 The scrambling towards a desired direction can be eliminated by a rule-based design of the TM parameters. 
It should be noted that the energy utilization efficiency of TMA reduces with  the periodic deactivation of antenna elements~\cite{hou2023energy}.

Recent research \cite{xu2025tmirs} shifts time modulation to an intelligent reflecting surface (IRS). IRS is a passive metasurface composed of programmable elements that dynamically adjust the phase of incident electromagnetic waves to realize beamforming \cite{wu2019towards}. By exerting the periodic TM on each IRS element, the system in \cite{xu2025tmirs} is designed to implement a 3D directional modulation. Also, the large aperture of an IRS delivers substantial beamforming gain that compensates for power lost of TMA during element deactivation. In \cite{xu2025tmirs}, the TM-IRS parameters are still obtained using simple, closed-form rules. Although these rule-based patterns are straightforward to implement, 
they ensure undistorted signal reception in only a single user direction, while their  extension to multiple users is challenging. This is particularly restrictive, as multi-user scenarios are common in contemporary wireless communication systems.

{This paper proposes a generative-AI framework for TM-IRS-assisted OFDM systems that replaces rule-based TM-IRS patterns 
with
selecting good sets of  TM-IRS parameters out of a set of all possible quantized ranges of parameters. The measure
of goodness is the sum rate at  authorized directions. 
We
model the TM-IRS parameter selection as a deterministic Markov
decision process (MDP), where each terminal state corresponds
to a specific configuration of TM-IRS parameters. GFlowNets
are employed to learn a stochastic policy that samples TM-IRS
parameter sets with probability proportional to their associated
sum rate reward.}
GFlowNets are unsupervised models that leverage a flow-matching principle to sample composite objects with probabilities proportional to a user-defined reward \cite{bengio2021flow, bengio2023gflownet}. In our setting, the analytical expression for the sum rate of the legitimate users serves as the reward function.
The TM-IRS parameter space is  discretized, and the parameter selection task is formulated as a deterministic Markov decision process (MDP), where each terminal state corresponds to a complete TM-IRS parameter set. i.e., a full on/off and phase shift configuration of the IRS elements. A feedforward neural network-based GFlowNet is trained offline to sample TM-IRS parameter sets in proportion to their associated sum rate reward. Experimental results show that the TM-IRS patterns generated by the learned GFlowNet yield undistorted reception at multiple desired directions while effectively scrambling signals in all other directions, thereby thwarting potential eavesdroppers. Notably, 
the sampling policy is stochastic and remains unknown to any adversary, significantly increasing the difficulty of potential attacks. Moreover, unlike deterministic optimization methods, the GFlowNet naturally generates a diverse set of high-performing TM-IRS configurations. This diversity allows the system to randomize TM-IRS patterns over time, thereby further enhancing the security \cite{tao2025twc}.

The remainder of the paper is organized as follows. Section~\ref{SystemModel} presents the system model and formalizes the TM-IRS parameter optimization problem. Section~\ref{Method} details the proposed GFlowNet-based TM-IRS parameter design framework. Numerical results demonstrating the effectiveness of the proposed approach are provided in Section~\ref{Experiments}. Finally, Section~\ref{Conclusion} concludes the paper and discusses potential directions for future research.




\section{System Model}\label{SystemModel}
Consider an IRS composed of $M_x \times M_z$ passive reflecting units that assists a uniform linear array (ULA) transmitting OFDM signals. Let $\theta_T$ and $\phi_T$ denote the elevation and azimuth angles of the ULA transmitter w.r.t. the IRS, respectively. From the transmitter's perspective, the IRS is modeled as a point target due to the sub-wavelength size of individual IRS elements and the overall compactness of the surface~\cite{hua2023secure}. Thus, we denote the direction of the IRS as viewed from the ULA by $\theta_I$. We consider a single legitimate user first for notational convenience and denote its direction relative to the IRS as $(\theta_c, \phi_c)$. It is assumed that $\theta_I$ is known at the transmitter, and $(\theta_T, \phi_T)$ and $(\theta_c, \phi_c)$ are known at the IRS. The location of potential eavesdroppers is assumed to be unknown. Additionally, all elements of both the ULA and the IRS are spaced by half the carrier wavelength, i.e., $\lambda/2$. {The channel to the legitimate destination is assumed to be known at the legitimate receivers. If the eavesdropper does not know its channel from the transmitter, then the channel effect would represent additional scrambling. Here we will assume the scenario in which the eavesdropper knows its channel and can compensate for it. Thus, in the following, the channel will not be included in the expressions.}

Each antenna element is fed with an OFDM signal, which is expressed as
\begin{equation}\label{eq1}
    e(t) = \frac{1}{\sqrt{K}} \sum_{k=0}^{K-1} d(k) e^{j2\pi(f_c + k f_s)t}, \quad 0 \leq t < T_s,
\end{equation}
where $K$ is the number of subcarriers, $d(k)$ is the digitally modulated data symbol on the $k$-th subcarrier, $f_c$ is the carrier frequency, $f_s$ is the subcarrier spacing, and $T_s$ is the OFDM symbol duration. By setting the antenna weight $w_n = e^{-j n \pi \cos \theta_I}$ to focus the ULA beam toward the IRS, we have the radiated waveform $r(t,\theta_I) = \frac{1}{\sqrt{N}} \sum_{n=0}^{N-1} e(t) w_n e^{j n \pi \cos \theta_I} = \sqrt{N} e(t)$, where $N$ is the number of transmit antennas.

Each IRS unit is connected to a high-speed SPST switch and a phase shifter. The switches operate in two states: ``on'' and ``off.'' Let $U_{mn}(t)$ denote the on/off switching function of the $(m,n)$-th IRS unit, with a period equal to $T_s$. Define the normalized turn-on instant as $\tau^o_{mn} \in [0,1)$ and the normalized on-duration as $\Delta\tau_{mn} \in [0,1)$. The switching function $U_{mn}(t)$ is set to 1 when $t \in [T_s \tau^o_{mn}, T_s (\tau^o_{mn} + \Delta\tau_{mn})]$ and 0 otherwise. This periodic square waveform can be expanded using its Fourier series:
\begin{equation}\label{eq2}
\begin{split}
    U_{mn}(t) = \sum_{l=-\infty}^{\infty} e^{j2\pi l f_s t} &\Delta\tau_{mn} \mathrm{sinc}(l\pi \Delta\tau_{mn}) \\
    &\times e^{-j l\pi (2\tau^o_{mn} + \Delta\tau_{mn})},
\end{split}
\end{equation}
The harmonics introduced by time modulation are centered at integer multiples of $f_s$. Define the far-field array factor of the $(m,n)$-th IRS element as\cite{yurduseven2020intelligent}
\begin{equation}
    a_{mn}(\theta, \phi) = e^{-j\pi(m \sin \theta \cos \phi + n \sin \theta \sin \phi)}.
\end{equation}
Let $c_{mn}$ be the unit-modulus phase shift applied by the $(m,n)$-th IRS unit, 
the signal reflected by the IRS toward direction $(\theta, \phi)$ can then be expressed as
\begin{equation}\label{eq3}
\begin{split}
    y(t, \theta, \phi) = r(t, \theta_I) \sum_{m=0}^{M_x-1} \sum_{n=0}^{M_z-1} &a_{mn}(\theta_T, \phi_T) U_{mn}(t) \\
    &\times c_{mn} a_{mn}(\theta, \phi).
\end{split}
\end{equation}

Substituting \eqref{eq1} and \eqref{eq2} into \eqref{eq3} and reorganizing terms yields
\begin{equation}
\begin{aligned}
    y(t, \theta, \phi) &= \sqrt{\frac{N}{K}} \sum_{k=0}^{K-1} d(k) e^{j2\pi(f_c + k f_s)t} \\
    &\quad \times \sum_{l=-\infty}^{\infty} e^{j2\pi l f_s t} V(l, \Omega_{mn}, \theta, \phi),
\end{aligned}
\end{equation}
where $\Omega_{mn} = \{c_{mn}, \Delta\tau_{mn}, \tau^o_{mn}\}$ represents the TM-IRS parameter configuration, and
\begin{equation}
\begin{aligned}
    V(l, \Omega_{mn}, \theta, \phi) &= \sum_{m=0}^{M_x-1} \sum_{n=0}^{M_z-1} a_{mn}(\theta_T, \phi_T) c_{mn} a_{mn}(\theta, \phi) \\
    &\quad \times \Delta\tau_{mn} \mathrm{sinc}(l\pi \Delta\tau_{mn}) e^{-j l\pi(2\tau^o_{mn} + \Delta\tau_{mn})}.
\end{aligned}
\end{equation}
Here, $V(l)$ denotes the coefficient of the $l$-th harmonic generated by the time modulation of the $(m,n)$-th IRS element at direction $(\theta, \phi)$. After OFDM demodulation and adding Gaussian noise $z_i$ with zero mean and variance $\sigma^2$, the received data symbol on the $i$-th subcarrier is given by
\begin{equation}\label{eq4}
    y_i(\theta, \phi) = \sqrt{\frac{N}{K}} \sum_{k=0}^{K-1} d(k) V(i-k, \Omega_{mn}, \theta, \phi) + z_i.
\end{equation}
From \eqref{eq4}, it is evident that each subcarrier symbol contains a weighted summation of symbols from all subcarriers, resulting in data scrambling across subcarriers.

In \cite{xu2025tmirs}, to ensure undistorted reception at the legitimate user, the TM parameters were selected to satisfy $V(i-k, \Omega_{mn}, \theta_c, \phi_c) = 0$ for all $i \neq k$, i.e. nulling scrambling, which can be achieved via closed-form rule-based TM-IRS parameter design. 
However, the resulting rules do not guarantee that the scrambling experienced by other legitimate users at $(\theta, \phi) \neq (\theta_c, \phi_c)$ can be effectively mitigated, making this approach difficult to extend to multi-user scenarios. In this work, we do not aim to enforce $V_{i-k} = 0$ for all $i \neq k$ (where $V_{i-k}$ denotes $V(i-k, \Omega_{mn}, \theta_c, \phi_c)$ for notational simplicity) to achieve undistorted reception. Instead, $V_{i-k}$ for $i \neq k$ can be treated as interference terms. Define the signal-to-interference-plus-noise ratio (SINR) at the $i$-th subcarrier of one legitimate user as
\begin{equation}
    \mathrm{SINR}_i = \frac{\eta |V_0|^2}{\eta \sum_{j=i-(K-1)}^{i} |V_j|^2 - |V_0|^2 + \sigma^2},
\end{equation}
where $\eta = N/K$. The achievable sum rate across all subcarriers can then be expressed as
\begin{equation}\label{sumC}
    C = \sum_{i=0}^{K-1} \log_2 (1 + \mathrm{SINR}_i).
\end{equation}
The total sum rate of $U$ legitimate users is
\begin{equation}\label{totalC}
    C_{\text{total}} = \sum_{u=1}^U C_u,
\end{equation}
where $C_u$ is defined by \eqref{sumC} and the subscript `$u$' denotes the $u$-th user. Meanwhile, to ensure that the symbol constellation remains unaffected by the phase of \( V_0 \), a constraint is imposed: \( |\arg(V_0)_u| \le \xi_u \), where \( \xi_u \) is a threshold determined by the modulation scheme. For \(\mathcal{M}\)-PSK modulation, \(\xi_u\) must be smaller than \( \pi/\mathcal{M} \). Accordingly, the design of the TM-IRS parameters is formulated as a constrained optimization problem:
\begin{equation}\label{optF}
\begin{split}
    &\max_{\Omega_{mn}} \quad C_{\text{total}} \\
    &\;\; \text{s.t.} \quad |\arg(V_0)_u| \le \xi_u.
\end{split}    
\end{equation}
From the above equations, we can see that the magnitude and phase of $V_0$, the scrambling terms, and the noise are all taken into account. In the following, we propose a GFlowNet-based approach to optimize the TM-IRS parameters under the constraint in~\eqref{optF}.


\section{GFlowNet-Based TM-IRS Parameter Design}\label{Method}

\subsection{Overview of GFlowNets}
The GFlowNet framework models the sequential decision-making process as a deterministic MDP, defined over a set of states $\mathcal{S}$, with a subset of terminal states $\mathcal{X} \subset \mathcal{S}$. At each state $s \in \mathcal{S}$, a discrete set of actions $\mathcal{A}(s)$ determines the permissible transitions, forming a directed acyclic graph (DAG) structure. A trajectory consists of a sequence of actions from the root (initial) state to a terminal state, with the possibility that different action paths may reach the same state, reflecting the non-injective structure of the graph. Rewards are only assigned to terminal states, while all intermediate states carry zero reward, i.e., $R(s) = 0$ for $s \notin \mathcal{X}$. The training objective in GFlowNets is to learn a stochastic policy that induces a distribution over terminal states proportional to their associated non-negative rewards \cite{bengio2021flow}.

To achieve this, GFlowNets view the MDP as a network of flows propagating from the root node to the terminal nodes. An edge flow $F(s, a)$ is defined for each action $a$ taken at state $s$, resulting in a transition to $s' = T(s,a)$, and the total state flow $F(s)$ corresponds to the sum of flows through that state. The flow matching principle requires that, at every state, the incoming flow equals the sum of its outgoing flow and reward. Specifically, for a node $s'$, we define the incoming and outgoing flows as:
\begin{equation}
    F_{\text{in}}(s') = \sum_{s,a : T(s,a) = s'} F(s,a),
\end{equation}
\begin{equation}
    F_{\text{out}}(s') = \sum_{a' \in \mathcal{A}(s')} F(s',a').
\end{equation}
Flow conservation imposes $F_{\text{in}}(s') = R(s') + F_{\text{out}}(s')$. From these flows, we define the forward and backward transition probabilities as
\begin{equation}
    P^F(s'|s) = \frac{F(s,a)}{F(s)}, \quad P^B(s|s') = \frac{F(s,a)}{F(s')},
\end{equation}
where $T(s,a) = s'$. The overall normalization constant, or partition function, of the flow network is given by the sum of rewards over all terminal states:
\begin{equation}
    Z = \sum_{x \in \mathcal{X}} R(x).
\end{equation}

To train the GFlowNet, the trajectory balance (TB) loss \cite{malkin2022trajectory} is used, which considers entire trajectories from the initial to terminal states. For a sampled trajectory $\tau = (s_0 \to s_1 \to \dots \to s_n = x)$, the TB objective compares the forward and backward path probabilities, scaled by the estimated reward and partition function:
\begin{equation}\label{TBL}
    L_{\mathbf{w}}(\tau) = \left( \ln \frac{Z_{\mathbf{w}} \prod_{t=1}^n P^F_{\mathbf{w}}(s_t|s_{t-1})}{R(x) \prod_{t=1}^n P^B_{\mathbf{w}}(s_{t-1}|s_t)} \right)^2,
\end{equation}
where both $P^F_{\mathbf{w}}$ and $P^B_{\mathbf{w}}$ are parametrized using neural networks with learnable parameters $\mathbf{w}$, and $Z_{\mathbf{w}}$ is a trainable scalar approximating the partition function. Minimizing this loss over sampled trajectories encourages the learned forward policy to produce samples whose marginal distribution over terminal states aligns proportionally with their rewards.

\subsection{GFlowNets for the TM-IRS Parameter Design}
We leverage the GFlowNet framework to optimize $\Omega_{mn}$ for all IRS elements in our OFDM system. The TM-IRS optimization is casted first as a parameter selection problem and a discrete MDP, where each intermediate state corresponds to a partial assignment of TM-IRS parameters. Specifically, each TM-IRS parameter, including $c_{mn}$, $\tau_{mn}^o$ and $\Delta\tau_{mn}$  for each IRS element, is discretized into $Q_1$, $Q_2$ and $Q_3$ possible values, i.e., 
$e^{j0}, e^{j\frac{2\pi}{Q_1}}, e^{j\frac{4\pi}{Q_1}}, \cdots, e^{j\frac{2\pi (Q-1)}{Q_1}}$ for $c_{mn}$, 
$0, \frac{1}{Q_2}, \frac{2}{Q_2}, \dots, \frac{Q_2-1}{Q_2}$ for $\tau_{mn}^o$, and $0, \frac{1}{Q_3}, \frac{2}{Q_3}, \dots, \frac{Q_3-1}{Q_3}$ for $\Delta\tau_{mn}$. 
Let $M = M_x M_z$ denote the total number of IRS elements. We represent the current TM-IRS state  by a binary vector $\boldsymbol{s} \in \mathbb{R}^{M \times Q}$, which is partitioned into $M$ blocks, each corresponding to one TM-IRS parameter and having $Q = Q_1 + Q_2 + Q_3$ entries.

Initially, at the root state, $\boldsymbol{s}$ is a zero  vector, meaning no any TM-IRS parameter has been assigned a value. After each action, a specific TM-IRS parameter is assigned one of its discretized values, by setting the corresponding entry in the associated block of $\boldsymbol{s}$ to 1 while keeping all other entries in that block at 0. After a sequence of $3M$ actions, a terminal state is reached where every TM-IRS parameter has been assigned exactly one value, and thus every block in $\boldsymbol{s}$ contains a single 1. At each step, the action space $\mathcal{A}(s)$ consists of choosing a value for one of the unassigned TM parameters. The reward associated with a terminal state is based on the objective in~\eqref{optF}, but modified to suit the GFlowNet framework. Specifically, we define the reward as
\begin{equation}\label{reward}
    R = C_{\text{total}} \cdot \prod_{u=1}^{U} \mathcal{H}(\xi_u - |\arg(V_0)_u|),
\end{equation}
where \( \mathcal{H}(\cdot) \) is the Heaviside step function, i.e., \( \mathcal{H}(x) = 1 \) if \( x \ge 0 \), and \( 0 \) otherwise. This formulation encourages the GFlowNet to generate TM-IRS parameter configurations that maximize the legitimate communication user performance only if the phase constraint \( |\arg(V_0)_u| \le \xi_u \) is satisfied for \emph{all} users. Infeasible solutions that violate any user’s constraint are assigned zero reward and are thus disincentivized during training. In the special case of a single legitimate user, we can directly set the phase shift parameter as \( c_{mn} = \left[a_{mn}(\theta_T, \phi_T)a_{mn}(\theta_c, \phi_c)\right]^{-1} \) to steer the IRS beam toward the legitimate direction. This choice ensures that \( V_0 \) becomes real-valued, thereby avoiding any undesired phase shift in the symbol constellation, and eliminates the need to enforce the constraint \( \prod_{u=1}^{U} \mathcal{H}(\xi_u - |\arg(V_0)_u|) \) in the reward.

The forward and backward sampling policies, $P_{\mathbf{w}}^F$ and $P_{\mathbf{w}}^B$, are modeled by a feedforward neural network parameterized by $\mathbf{w}$. The output of the network is a vector of dimension $2M \times Q$, where the first $M \times Q$ entries correspond to the forward transition probabilities and the latter $M \times Q$ entries correspond to the backward transition probabilities. During training, the action selection is based on the forward probabilities $P_{\mathbf{w}}^F$. To prevent repeated selection of already assigned parameters, the forward probabilities for completed parameters are masked to zero at each decision step. The network is trained using the TB loss described in \eqref{TBL}, ensuring that the learned forward policy samples TM-IRS parameter configurations with probability proportional to their associated reward in \eqref{reward}. Training is conducted offline by sampling multiple root-to-leaf trajectories in the MDP, applying the TB loss, and updating $\mathbf{w}$ and the total reward $Z$ via gradient descent. After training, the GFlowNet can be deployed online to sample diverse high-reward TM-IRS parameter configurations.


\section{EXPERIMENTS}\label{Experiments}

\begin{figure}[t]
\centerline{\includegraphics[width=3.4in]{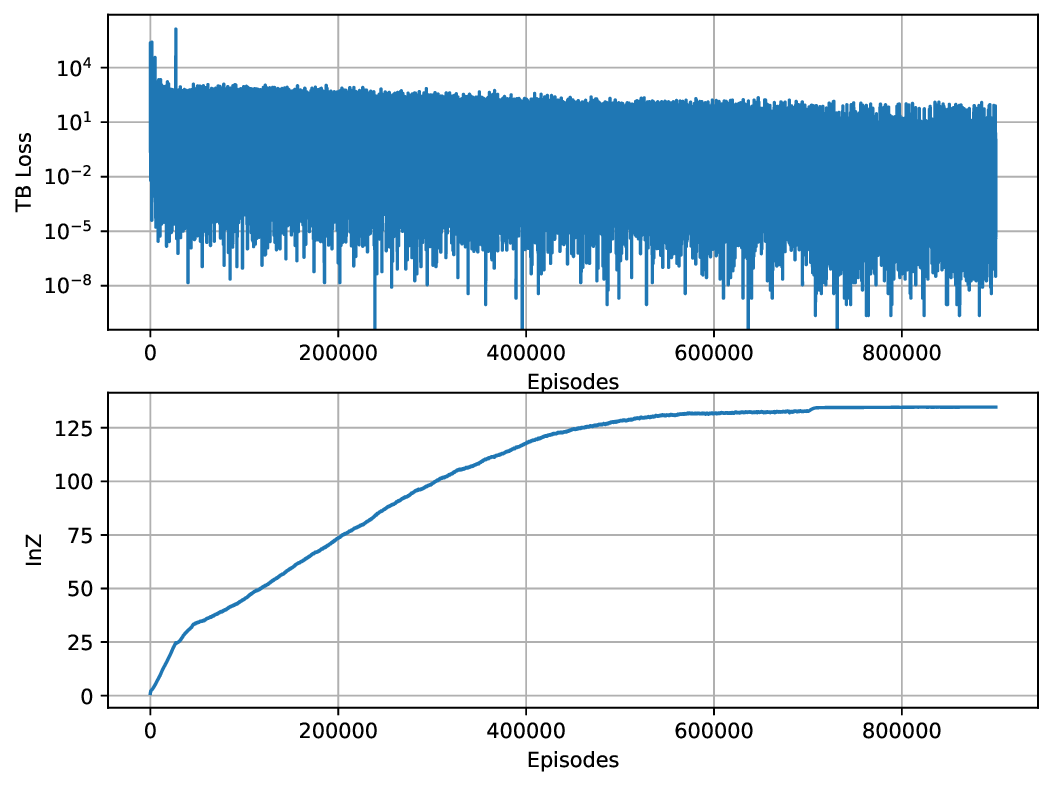}}
\caption{Evolution of the TB loss and the estimated partition function $\ln Z$ over training episodes.}
\label{fig1}
\end{figure}


We consider an IRS-assisted OFDM system with $M_x = M_z = 6$ passive reflecting elements, $K = 16$ subcarriers, transmitting 1024 OFDM signals, and $N = 8$ antennas at the ULA transmitter. The transmitter is located at $(\theta_T, \phi_T) = (15^\circ, 10^\circ)$. QPSK modulation is employed, the SNR is set to 0 dB and the path loss is set as 1 for simplicity. Also, we adopt a nearest-neighbor decision rule to detect the transmitted symbol. For the adopted GFlowNet, $Q_1$ is set as 16 and each TM parameter, $\tau_{mn}^o$ and $\Delta \tau_{mn}$, is discretized into $Q_2 = Q_3 = 8$ uniform values in $[0,1)$ unless otherwise specified. A feedforward neural network with three hidden layers, each containing 256 neurons, is used to parameterize the GFlowNet. Training is performed offline on a powerful Apple M3 Max chip with 36 GB memory.

Here, symbol error rate (SER) is adopted as the performance metric. If evaluated on a logarithmic scale, a small offset of $10^{-4}$ is added to handle zero-SER cases if necessary. In the SER heatmaps, darker regions indicate lower error rates.

\begin{figure}[t]
    \centering

    \begin{minipage}{0.235\textwidth}
        \centering
        \includegraphics[width=\textwidth]{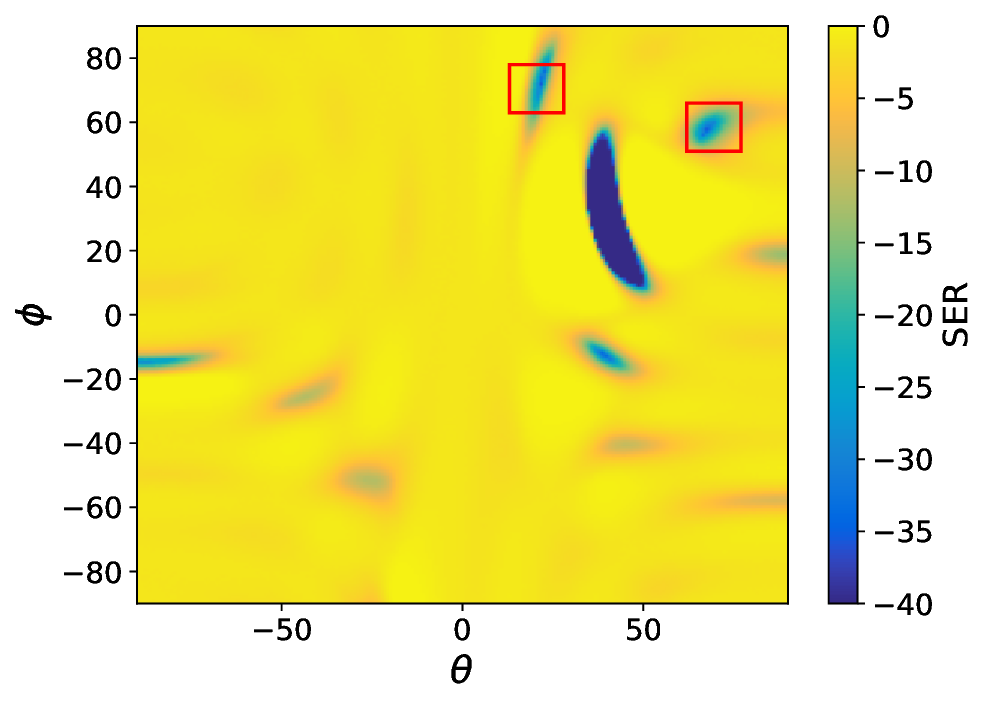}
        \vspace{0.5em}  
        \small (a)
    \end{minipage}
    \begin{minipage}{0.235\textwidth}
        \centering
        \includegraphics[width=\textwidth]{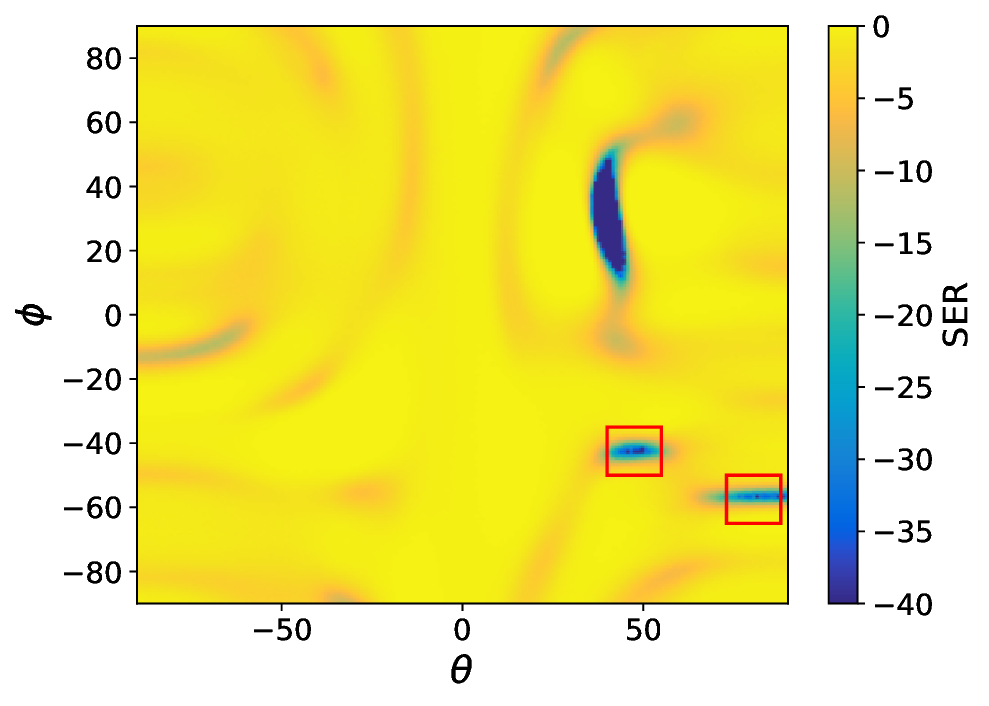}
        \vspace{0.5em}
        \small (b)
    \end{minipage}

    \caption{Comparison of SER over different spatial directions: (a) rule-based TM parameter design~\cite{xu2025tmirs}; (b) GFlowNet-based TM parameter design.}
    
    \label{fig2}
\end{figure}

\begin{figure}[t]
    \centering

    \begin{minipage}{0.235\textwidth}
        \centering
        \includegraphics[width=\textwidth]{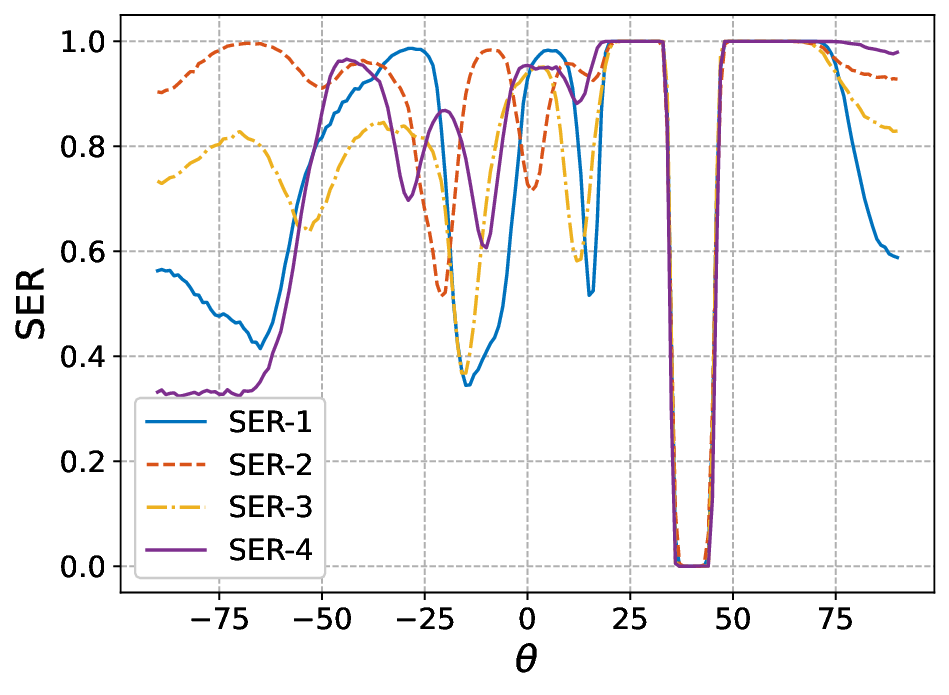}
        \vspace{0.5em}  
        \small (a)
    \end{minipage}
    \begin{minipage}{0.235\textwidth}
        \centering
        \includegraphics[width=\textwidth]{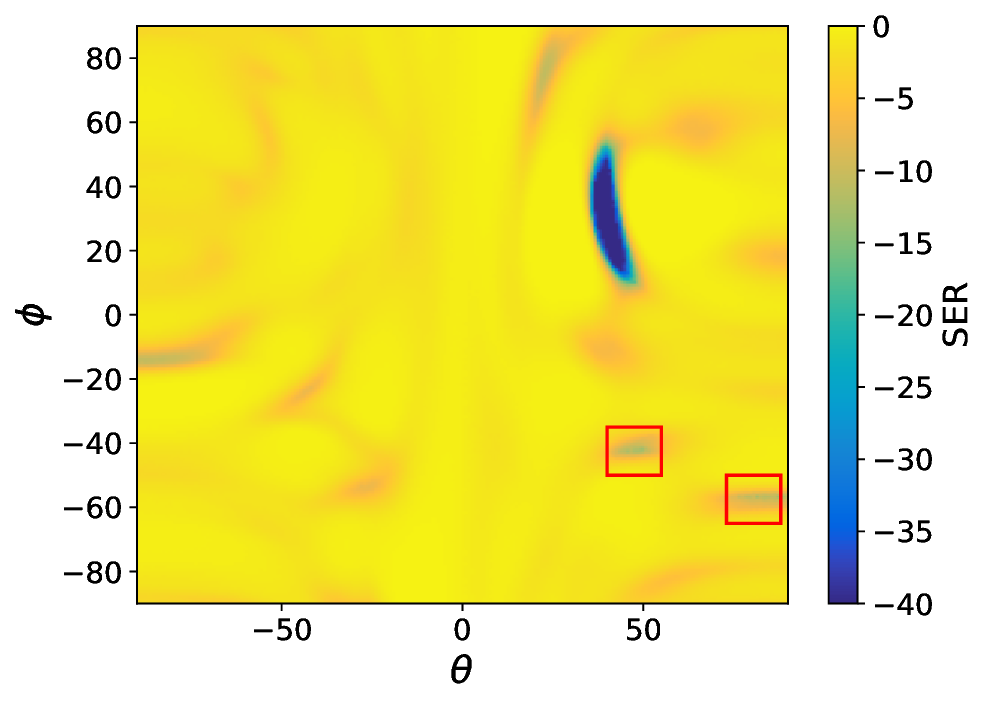}
        \vspace{0.5em}
        \small (b)
    \end{minipage}

    \caption{Enhancing security via TM parameter diversity: (a) SER versus $\theta$ for four GFlowNet-generated TM configurations with fixed $\phi=30^\circ$; (b) averaged SER across the four configurations.}
    \label{fig3}
\end{figure}

We begin with a single legitimate user located at \( (\theta_c, \phi_c) = (40^\circ, 30^\circ) \) to efficiently demonstrate the performance of the proposed GFlowNet-based design and to facilitate a fair comparison with the rule-based TM approach in~\cite{xu2025tmirs}. Here \( c_{mn} = \left[a_{mn}(\theta_T, \phi_T)a_{mn}(\theta_c, \phi_c)\right]^{-1} \), so $c_{mn}$ is not included in the GFlowNet and the training time can be reduced greatly. Also, the GFlowNet model is trained using $9 \times 10^5$ sampled trajectories, with a learning rate of $10^{-2}$ for the first $7 \times 10^5$ trajectories to accelerate the training and $10^{-3}$ for the remaining $2 \times 10^5$ to fine-tune the training.

Fig.~\ref{fig1} shows the evolution of the TB loss and the estimated partition function $\ln Z$ over training episodes. The TB loss steadily decreases, indicating that the forward and backward flows are being balanced properly. The partition function $\ln Z$ stabilizes as training progresses, suggesting convergence of the overall model. It is worth noting that the TM parameter space contains approximately $8^{72} \approx 10^{65}$ configurations, making exhaustive search infeasible. However, by parametrizing the flow using a deep neural network, the proposed framework effectively generalizes across the enormous solution space using only $9 \times 10^5$ samples (fewer than 0.000001\% of all possible configurations), inferring reward distribution even for unvisited TM configurations.


{Fig.~\ref{fig2} compares the SER performance across spatial directions for two TM design methods: the rule-based approach from~\cite{xu2025tmirs} in Fig.~\ref{fig2}(a), and the proposed GFlowNet-based method in Fig.~\ref{fig2}(b). In both cases, the desired user direction \( (40^\circ, 30^\circ) \) achieves very low SER, while most undesired directions exhibit high SER, indicating that the proposed method can achieve comparable direction modulation performance for security against the rule-based one. Moreover, several unintended directions also experience low SER, as highlighted by the red boxes in Fig.~\ref{fig2}(b). This arises because our proposed method does not explicitly regulate the SINR in undesired directions; as a result, certain TM-IRS configurations may inadvertently yield high SINR in those regions. To mitigate this situation, we can leverage the GFlowNet's capability to generate diverse high-reward TM configurations and vary the TM pattern over time. Specifically, four distinct TM parameter sets are sampled, and the configuration is switched every 256 OFDM symbols. Fig.~\ref{fig3}(a) illustrates the SER versus \( \theta \) (with fixed \( \phi = 30^\circ \)) for each of the four configurations individually. It can be seen that low-SER directions differ across configurations, while the desired user direction consistently maintains near-zero SER. Fig.~\ref{fig3}(b) shows the aggregated SER performance across all spatial directions, where the SER in previously vulnerable regions is  improved, as evidenced by the lighter color areas. This dynamic TM strategy effectively reduces the risk of eavesdropping without requiring knowledge of the eavesdroppers’ locations.}

\begin{figure}[t]
\centerline{\includegraphics[width=3.0in]{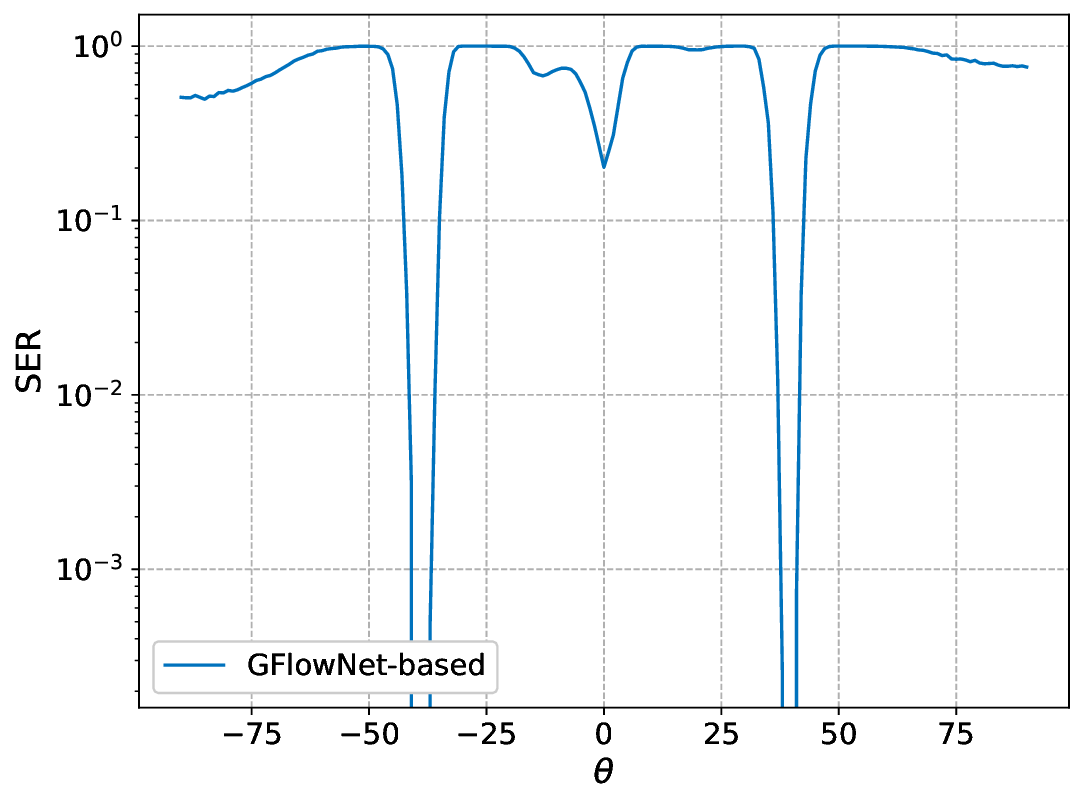}}
\caption{SER versus \( \theta \) for the proposed GFlowNet-based method in a two-user scenario.
}
\label{fig6}
\end{figure}

To evaluate the multi-user capability of the proposed TM-IRS design, we extend the previous single-user setting to a two-user scenario by introducing a second legitimate user located at \( (-40^\circ, 30^\circ) \), in addition to the original user at \( (40^\circ, 30^\circ) \). In this case, the SNR is set as 0 dB and the GFlowNet is trained using the  reward formulation in~\eqref{reward}, where $\xi_u$ is set as $\pi/5$ for both users. 
Fig.~\ref{fig6} presents the SER performance versus angle \( \theta \) with fixed $\phi=30^\circ$, from which we can see that the GFlowNet-based approach successfully maintains low SER for both user directions, thereby demonstrating its ability to jointly optimize TM-IRS parameters for multi-user support. This highlights the flexibility and scalability of the proposed framework in handling more complex directional modulation requirements.



\section{CONCLUSION}\label{Conclusion}
This paper presented a new framework for 3D directional modulation in OFDM systems assisted by TM-IRS. Motivated by the limitations of existing rule-based TM-IRS designs—which are typically restricted to single-user scenarios and lack flexibility—we proposed a generative model based on GFlowNets to enable secure multi-user transmission. By modeling the TM-IRS parameter selection process as a deterministic MDP, and training a trajectory balance-based GFlowNet to sample from this space, we were able to generate diverse TM-IRS configurations that maximize the achievable sum rate across authorized directions while maintaining constellation integrity. Experimental results confirmed that the proposed method offers strong performance in both single- and multi-user settings, demonstrating lower SERs in desired directions while introducing sufficient scrambling elsewhere. Also, the generated diverse high-performing parameter sets facilitate time-varying TM-IRS strategies that further enhance security against potential eavesdroppers even without prior knowledge of their locations. Future work may explore extensions to more complex wireless scenarios, such as (partially) known locations of the eavesdroppers, imperfect CSI, or joint optimization with sensing functions.


\bibliography{Paper2025}
\bibliographystyle{IEEEtran}

\end{document}